\documentstyle[11pt,moriond,epsfig,amssymb]{article}

\def\Journal#1#2#3#4{{#1} {\bf #2}, #3 (#4)}

\def\PRL{\em Phys. Rev. Lett.}

\def\PRC{{\em Phys. Rev.} C}

\def\be{\begin{equation}}
\def\ee{\end{equation}}
\def\bea{\begin{eqnarray}}
\def\eea{\end{eqnarray}}

\begin{document}
\vspace*{4cm}

\title{Review on new Neutrino Oscillation Experiments}

\author{Mario Campanelli}

\address{Institut f\"ur Teilchenphysik ETHZ, CH-8093 Z\"urich, Switzerland}

\maketitle\abstract{Driven by new experimental results, in the latest 
period several new neutrino oscillation experiments have been proposed.
I will outline the main ideas behind the different proposals, in particular
concerning atmospheric neutrinos and neutrinos from accelerated beams.
}




\newpage
\section{Introduction}
Neutrino physics is living a period of great excitement. Many experiment 
show results that point towards evidence for neutrino oscillations,
and others are planned to verify with better precision or new techniques
these results. In particular, the experiments observing effects difficult
to explain without oscillations can be classified into three categories:
\begin{itemize}
\item solar neutrinos \par
A deficit of $\nu_e$ is observed the solar neutrino spectrum. It can be
interpreted as oscillations with $\Delta m^2\approx 10^{-5} eV^2$ (if
oscillation mainly takes place in the solar matter) or 
$10^{-9}-10^{-10} eV^2$ (if  oscillation takes place in the vacuum between
the sun and the earth)
\item atmospheric neutrinos \par
the $\nu_\mu/\nu_e$ ratio is lower in the data than in the expectations;
it can be interpreted in a disappearance of $\nu_\mu$, governed by 
oscillations with $\Delta m^2\approx 10^{-3}-10^{-2} eV^2$
\item LSND \par
the LSND experiment at Los Alamos observes an excess of electrons in a
$\nu_\mu$ beam. If interpreted as neutrino oscillation, the mass difference
would be $\Delta m^2\approx 1 eV^2$
\end{itemize}

Three neutrino families have only two possible mass difference scales, so
it is difficult to accommodate the present experimental data without either
stretching them, discarding one result, or assuming new phenomena, like the
existence of sterile neutrinos not coupling to the Z.\par
\section{Atmospheric neutrino results}
Neutrinos are produced in the atmosphere from $\pi^\pm\to\mu^\pm\nu_\mu$
and subsequent $\mu^\pm\to e^\pm\nu_e\nu_\mu$ decays. The calculations
of absolute fluxes have uncertainties of the order of 30\%, so what is
usually used is the quantity $R=\frac{\Phi_{\nu_e+\bar{\nu}_e}}
{\Phi_{\nu_\mu+\bar{\nu}_\mu}}\approx \frac{1}{2}$ where most of the
uncertainties cancel out, leading to an error of about 5\%.\par
The atmospheric neutrino experiments are detecting electrons or muons 
produced in charged-current neutrino interactions, using calorimeters or
water Cherenkov devices. The results obtained for the double ratio by the
different experiments, are shown in table \ref{tab:atmres}. The statistical
accuracy of the latest measurements of the SuperKamiokande \cite{sk}
collaboration is extremely good, given the large mass of the detector,
so the statistical error is smaller than the systematics due to the 
uncertainties on the double ratio, and even with more data no further
improvement is foreseen.\par
\begin{figure}[tbh]
  \begin{center}
   \includegraphics[width=8cm,bb= 73 130 570 710,clip]{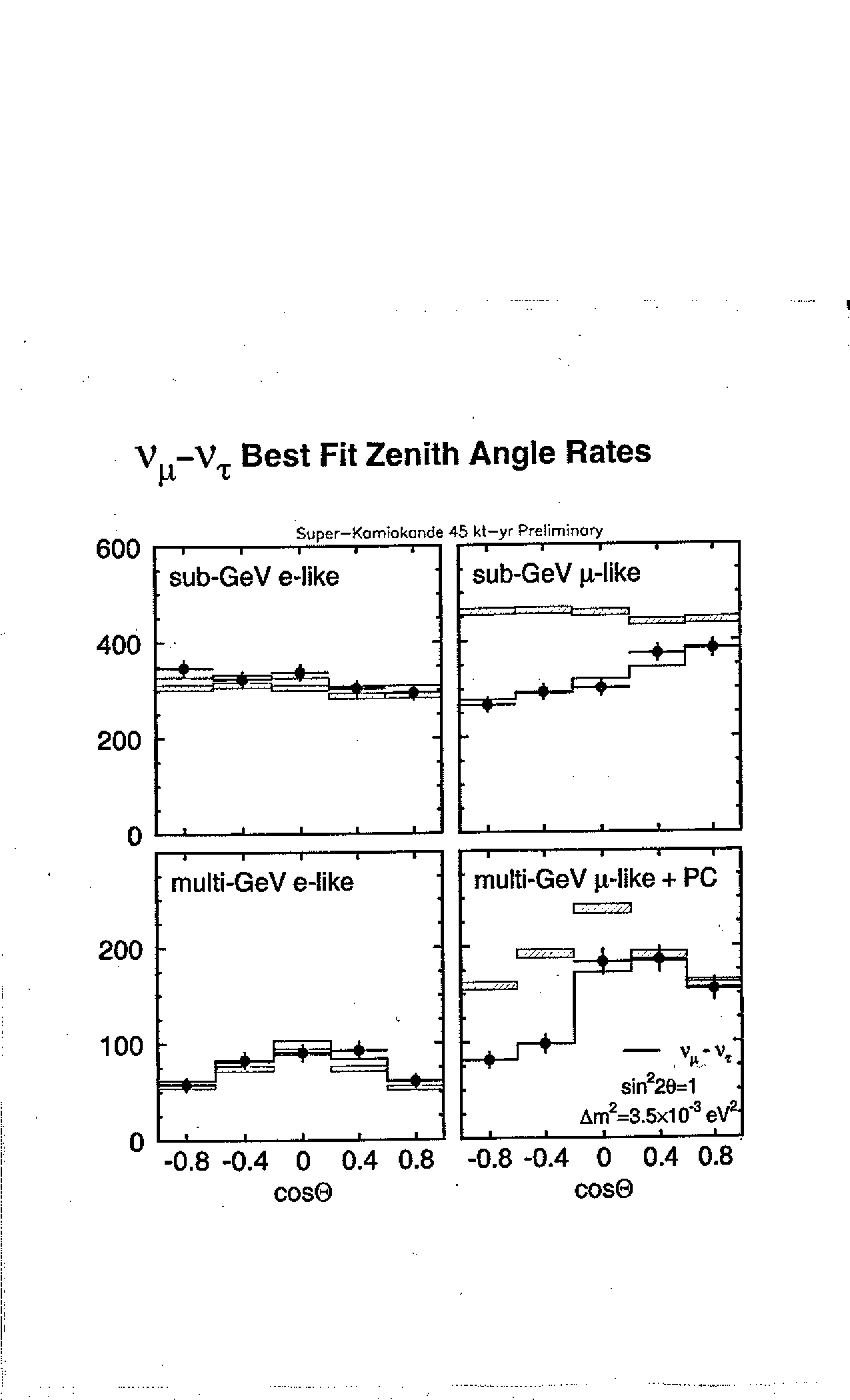}
  \end{center}
\label{fig:azimsk}
\caption{Azimuthal distribution of neutrio events in SuperKamiokande.
Dots with error bars correspond to data, boxes to MonteCarlo with no 
oscillation, histogram to MonteCarlo with $\nu_\mu\to\nu_\tau$ oscillations
with $\Delta m^2=3.5\times 10^{-3}$.}
\end{figure}

\begin{table}[tbh]
\begin{center}
\begin{tabular}{lll}\hline\hline
{\bf Experiment} & {\bf Exposure} & {\bf \hspace*{3.cm}R} \\
 & {\bf (Kton-year)}  & \\ \hline
SuperK subGeV & \hspace*{1.cm}45 &  $0.67 \pm 0.02 \pm 0.0
5$ \\
SuperK multiGeV & \hspace*{1.cm}45 & $0.66 \pm 0.04 \pm 0
.08$ \\
IMB & \hspace*{1.cm}7.7 & $0.54 \pm 0.05 \pm 0.11$ \\
Kamiokande subGeV & \hspace*{1.cm}7.7 & $0.60 \pm 0.06 \pm 0.07$ \\
Kamiokande multiGeV & \hspace*{1.cm}7.7 & $0.57 \pm 0.08 \pm 0.08$ \\
Soudan-II & \hspace*{1.cm}4.2 & $0.66 \pm 0.11 \pm 0.06$ \\
NUSEX & \hspace*{1.cm}0.4 & $0.96 + 0.32 - 0.28$ \\
Fr\'ejus & \hspace*{1.cm}2.0 & $1.00 \pm 0.15 \pm 0.08$ \\
\end{tabular}
\label{tab:atmres}
\caption{Double ratio $(N_\mu/N_e)_{Data}/(N_\mu/N_e)_{MonteCarlo}$ for
different atmospheric neutrino detectors. The first five values refer to
water Cherenkov detectors, the last three to calorimeters. Of the errors
quoted, the first is statistical and the second systematical.}
\end{center}
\end{table}
Before reaching a detector paced close to the surface, neutrinos produced 
in the atmosphere can travel distances ranging from about 10 km (neutrinos
coming from above) to about 10000 km (neutrinos coming from below, that
crossed the whole earth diameter before reaching the detector). Therefore,
a measurement of the zenith angle of the neutrino direction can be converted
into a measurement of the distance L between neutrino production and
detection.
Since the oscillation probability depends on the ratio $L/E$, if the
oscillation hypothesis is correct the $\nu_\mu$ deficit should not be
uniformly distributed in azimuthal angle, but in the case of small
probability be larger for neutrinos coming from below and smaller for
those coming from above. The fact that this effect is actually observed
by the SuperKamiokande collaboration (see figure \ref{fig:azimsk}), 
supports the oscillation hypothesis.
The disappearance of muon neutrino observed in the SuperKamiokande
experiment is therefore well-established, and it is likely to be due to
oscillations. In case it is interpreted as an indication of 
$\nu_\mu\to\nu_\tau$ oscillation, the best fit of the observed distribution
yields a value of $\Delta m^2$ between $1.5\times 10^{-3}$ and $8\times
10^{-3} eV^2$ at 90\% C.L., with mixing angle close to maximum. The$\chi^2$
from this fit is
$\chi^2_{\nu_\mu\to\nu_\tau}=62.1/67 DOF (P=65\%).$\par
Other interpretations are possible:\begin{itemize}
\item $\nu_\mu\to\nu_e$\par 
this possibility is ruled out by the result of the Chooz experiment looking
for $\nu_e$ disappearance; the ratio observed/expected events is
$0.98\pm0.04\pm0.04$, thus excluding $\nu_e\to\nu_x$ in the case of
maximal mixing for $\Delta m^2>9\times 10^4 eV^2$. Also the fit of the
Superkamiokande distribution seems disfavoring this hypothesis:
$\chi^2_{\nu_\mu\to\nu_e}=110/67 DOF (P<0.1\%)$.
\item $\nu_\mu\to\nu_s$\par
The SuperKamiokande fit leaves this possibility open ($\chi^2_{\nu_\mu\to
\nu_s}=64.3/67 DOF (P=57\%)$. More experimental indications can come from
the ratio $\pi^0/e$; in fact, the Neutral Current process $\nu_{e,\mu,\tau}
N\to\nu_{e,\mu,\tau}\pi^0 X$ does not exist for sterile neutrinos, since
they do not couple to the $Z^0$, modifying the ratio above. The 
SuperKamiokande result $(\pi^0/e)_{data}/(\pi^0/e)_{MC}= 0.93\pm 0.07 
(stat) \pm 0.19 (syst)$ disfavors this interpretation.
\end{itemize}
\section{Next Generation}
\begin{figure}[tbh]
  \begin{center}
   \includegraphics[width=10cm,bb=75 272 535 630,clip]{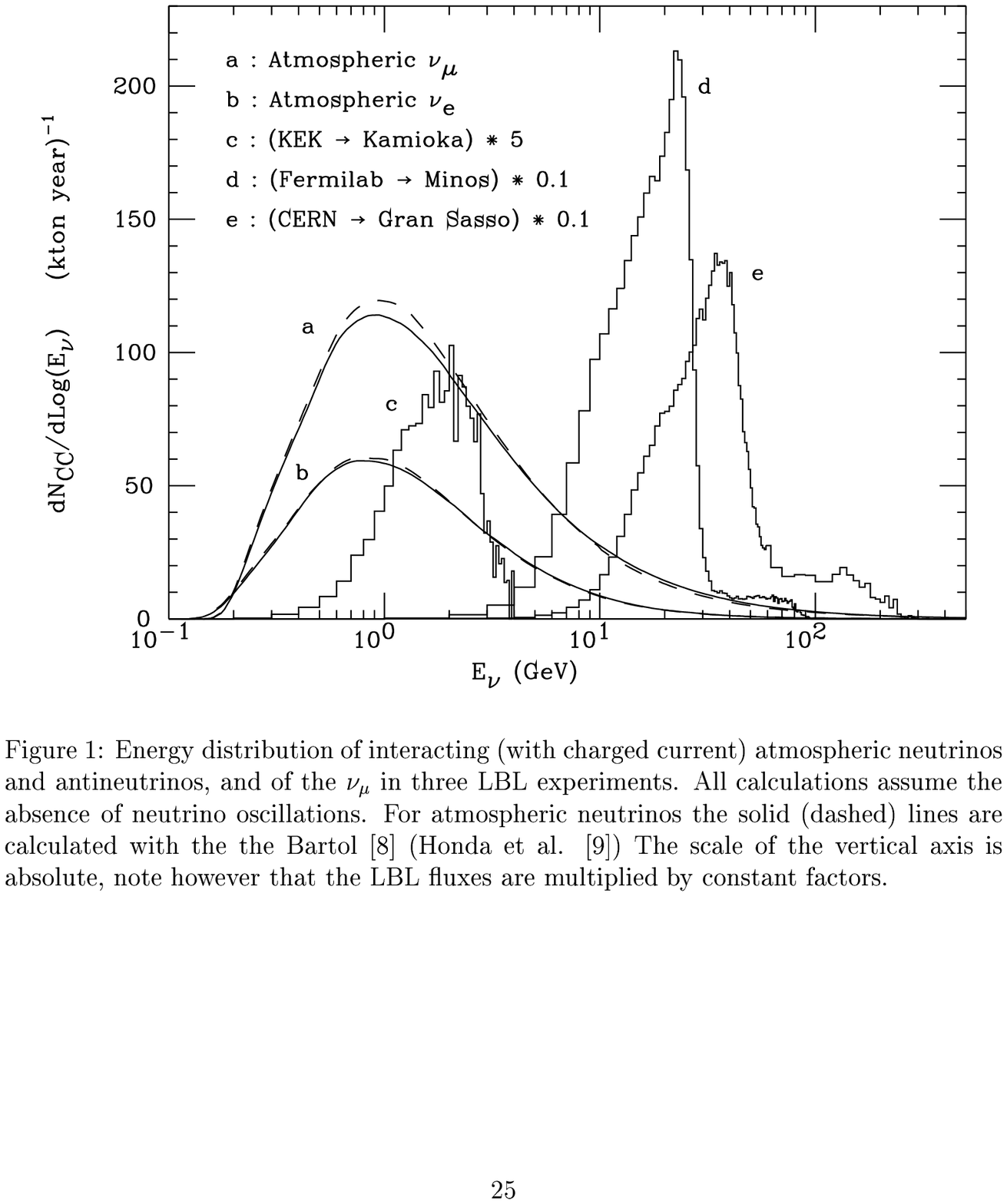}
  \end{center}
\label{fig:nuspec}
\caption{Energy spectra for the two flavours of atmospheric neutrinos,
as well as for the proposed long-baseline beams.}
\end{figure}

The oscillation hypothesis for atmospheric neutrinos has to be confirmed
and further investigated. The proposed future experiments can be split
into two main categories: new detectors for atmospheric neutrino studies,
or experiments performed with high-energy neutrino beams from accelerators.
The issue can be better understood looking at the neutrino energy spectra
for the different options (figure \ref{fig:nuspec}).\par
Atmospheric neutrinos have a broad spectrum (lines a and b), with their
maxima around 1 GeV. Only the tail of these events reaches energies
higher than the energy threshold for $\tau$ production (about 10 GeV).
Therefore, a next generation atmospheric neutrino experiment will aim
at verifying the results of SuperKamiokande using a different technique.
Artificial neutrino beams, on the other hand, are designed to have most
of the flux at the higher energy; the primary goal of experiments with
these beams will be $\tau$ lepton identification.
\subsection{Icarus at Gran Sasso}
The Icarus \cite{ica} detector is the only next-generation atmospheric
experiment currently approved. It consists in a liquid-argon TPC, with
3D imaging capabilities and very good particle identification. 
Given its density (1.4 $g/cm^3$) particles produce electromagnetic and
hadronic showers, and, as well as a tracking device, the detector can be
used as a calorimeter, with energy resolution $\sigma(E)/E=1\%+3\%/
\sqrt{E(GeV)}$ for electromagnetic showers, and $\sigma(E)/E=15\%/
\sqrt{E(GeV)}$ for fully contained hadronic showers, assuming a software
compensation is applied.\par
\begin{figure}[p]
  \begin{center}
   \includegraphics[width=10cm,bb=100 230 520 680,clip]{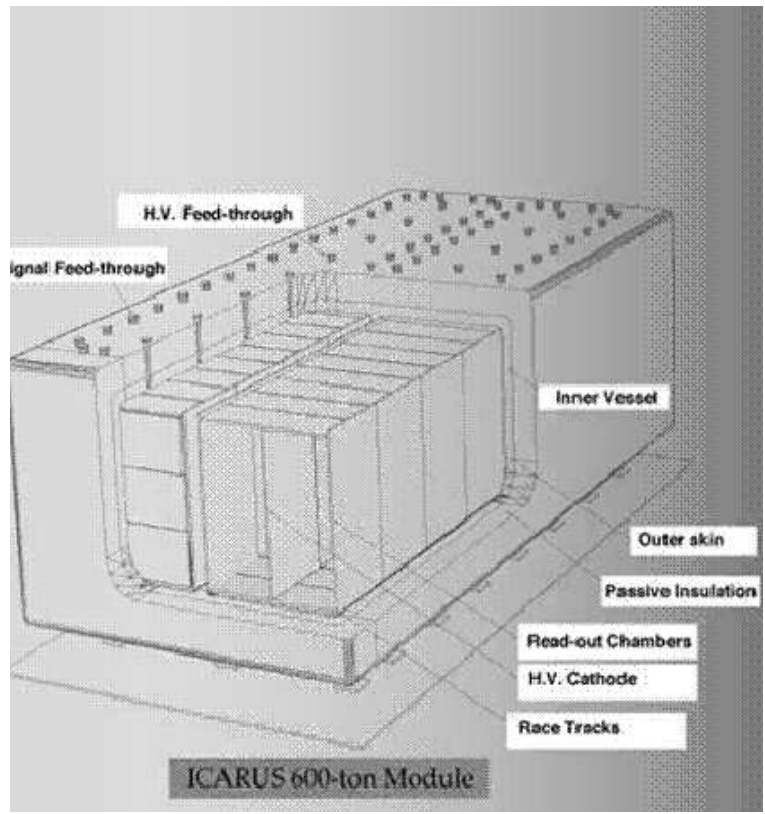}
  \end{center}
\label{fig:icar}
\caption{Schematic view of a 600 ton module of the ICARUS detector}
\end{figure}
\begin{figure}[p]
  \begin{center}
   \includegraphics[width=10cm]{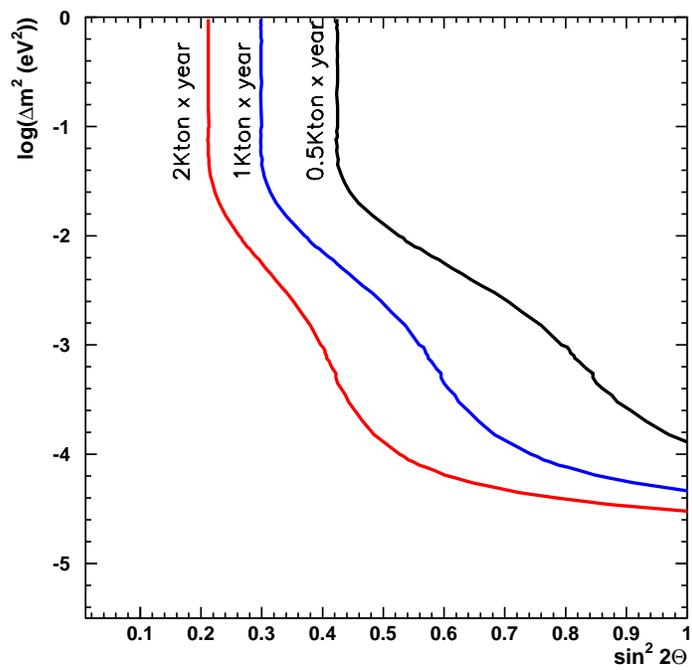}
  \end{center}
\label{fig:iclimit}
\caption{90\% C.L. intervals for $\nu_\mu$ disappearance from atmospheric
events for different exposures of the Icarus detector.}
\end{figure}
A 50 liter prototype of this detector was tested with neutrino of the
WANF beam at CERN in years 1997 and 1998. Several neutrino interaction
events have been recorded and studied, and an electron lifetime larger
than 8 ms was obtained. Given the fact that the drift velocity for an
electric field of 200 V/m is $1 mm/\mu s$, this step was very encouraging
towards the construction of a larger detector. Presently, a 15 ton module
is in operation, and the first 600 ton detector is under construction.
This detector is expected to start data taking in Gran Sasso in the 
beginning of year 2001, and an upgrade to 2.4 ktons is foreseen in the next
5 years.\par
Given the rate for atmospheric neutrinos of 100 events/0.5 kton/year,
after 1 year of data taking the whole SuperKamiokande suggested region
can be covered, giving the first independent test of this result.
It can look surprising that a small number of events can reach similar
sensitivity with respect to the much larger statistics of SuperKamiokande;
it is possible due to the better understanding of the final state and to
the lower energy threshold, allowing to explore a smaller value of
$\Delta m^2$.
\subsection{Large atmospheric experiments}
Future large atmospheric neutrino experiments are planned, in order to
collect more statistics, in particular in the high-energy region where
the uncertainties on neutrino production are smaller. New detection
techniques are also proposed; the experiments can be classified into
four categories:\begin{itemize}
\item high density iron calorimeter (CERN-SPSC 98-28)
\item lower density calorimeter with magnetic field (NICE)
\item large ICARUS-like detector (SUPER-I)
\item large water Cherenkov detector (AQUARICH)
\end{itemize}
\par
\begin{figure}[tbh]
  \begin{center}
   \includegraphics[width=10cm,bb=123 526 500 700,clip]{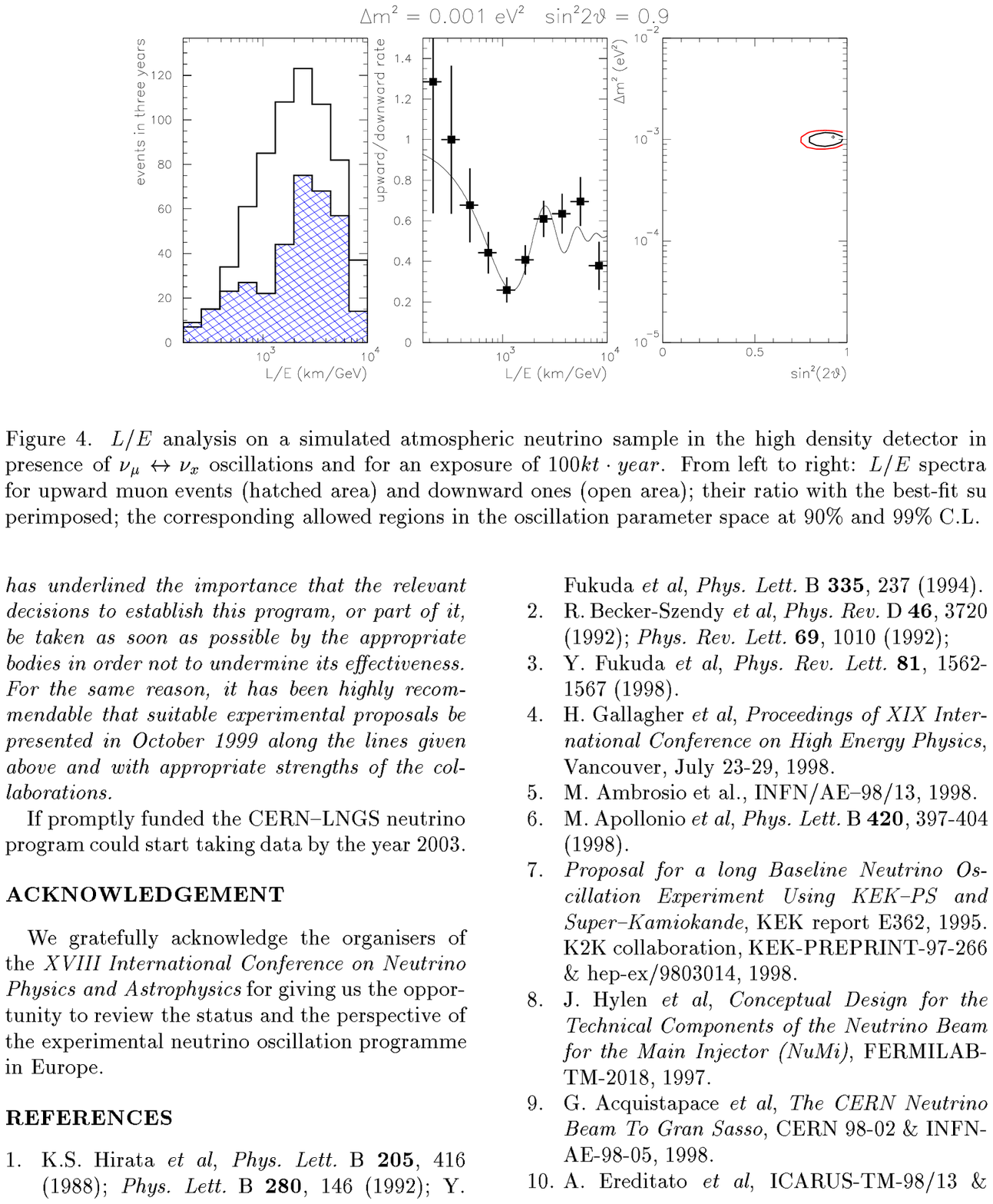}
  \end{center}
\label{fig:picchio}
\caption{First plot: downward-going (hatched histogram) and upward-going
(open histogram) neutrino events for $\nu_\mu\to\nu_\tau$ oscillations
with $\Delta m^2=1.\times 10^{-3}$. Second plot: ratio of the above 
distributions, showing a clear dip in correspondence with the first
maximum of oscillation probability. Third plot: accuracy of the measurement
of the oscillation parameters after three years.}
\end{figure}

A high-density calorimeter could be built using quite standard technologies
\cite{picchiom}. A large mass (i.e. 36 ktons of iron/tracker sandwich)
is needed to fully contain muons from $\nu_\mu$ interactions, and so measure
their energy from the range, and their direction, which for energetic
events is a good approximation of the direction of the incoming neutrino.
For values of $\Delta m^2<10^{-3}eV^2$, the downward-going muons do not
oscillate, and can be considered as the reference sample; upward-going
muons on the other hand do oscillate; the ratio of the $L/E_\nu$ 
distributions for the two samples (figure \ref{fig:picchio}) shows a dip 
in correspondence of the first maximum of the oscillation probability.
From the position of this dip a direct measurement of $\Delta m^2$
can be performed, while the depth allows a determination of the mixing
angle, determining the oscillation parameters with good precision.\par
In 4 years of data taking, such a detector could be sensitive to 
$\nu_\mu\to\nu_x$ with $\Delta m^2>6\times 10^{-5} eV^2$. Given these
characteristics, this approach is good only if the oscillation is governed
by small values of $\Delta m^2$; for values of this parameters larger than
$10^{-3} eV^2$ almost horizontal neutrinos should be used, introducing large
uncertainties; for even larger values, even the downward-going neutrinos
oscillate, and cannot be considered any more a reference sample.
Moreover, this kind of detector is not able to identify electron, and
very poor in reconstructing the hadronic energy, restricting the events
studied to a small fraction (about 10\%) of the total.\par
A slightly different approach is using a more granular calorimeter with 
smaller mass (about 10 kton), surrounding it by a magnetized iron 
spectrometer, as proposed by the NICE \cite{nice} group. Muon momenta can 
be then measured with the spectrometer, without the need of full 
containment; the mass can be smaller and the granularity improved, to
improve the neutrino direction reconstruction trough the measurement of
the hadronic energy.\par
If more granularity should be pursued, a possibility would be to build
a large detector using the ICARUS technology, and recently a proposal for 
a 30 kton SUPER-ICARUS \cite{superi} has been made. This detector would
have no magnetic field, but its good electron identification would open
much more study opportunity than the iron calorimeters.\par
The only technique to have very large detectors ($>100$ kton) at an 
affordable cost is probably the detection of Cherenkov light either in
water or in ice. The Aquarich group\cite{aquarich} proposed the construction
of a large Ring-Imaging Cherenkov made of two concentric spheres, the inner
one used as a reflecting surface, and the outer one equipped with
photomultipliers to catch the light produced by the Cherenkov rings.
Momentum and mass of the particles produced in the neutrino interaction are
measured from the width and opening angle of the rings, and a precision
$\sigma(p)/p=O(1\%)$ could be achieved. Presently prototype tests of this
detector are in progress.\par
\section{Long Baseline neutrino beams}
The main reasons to use neutrinos from accelerators is to have more control
on the neutrino flux and composition, and to have sufficient energy
to perform $\tau$ appearance experiments. Three programs are presently
competing:\begin{itemize}
\item K2K from KEK to Kamioka (235 Km, 1999-)
\item NuMi from FNAL to Soudan (734 Km, 2002-)
\item NGS from CERN to Gran Sasso (732 Km,2004-)
\end{itemize}
\subsection{K2K}
\begin{figure}[tbh]
  \begin{center}
   \includegraphics[width=5cm,bb=290 124 600 640,clip]{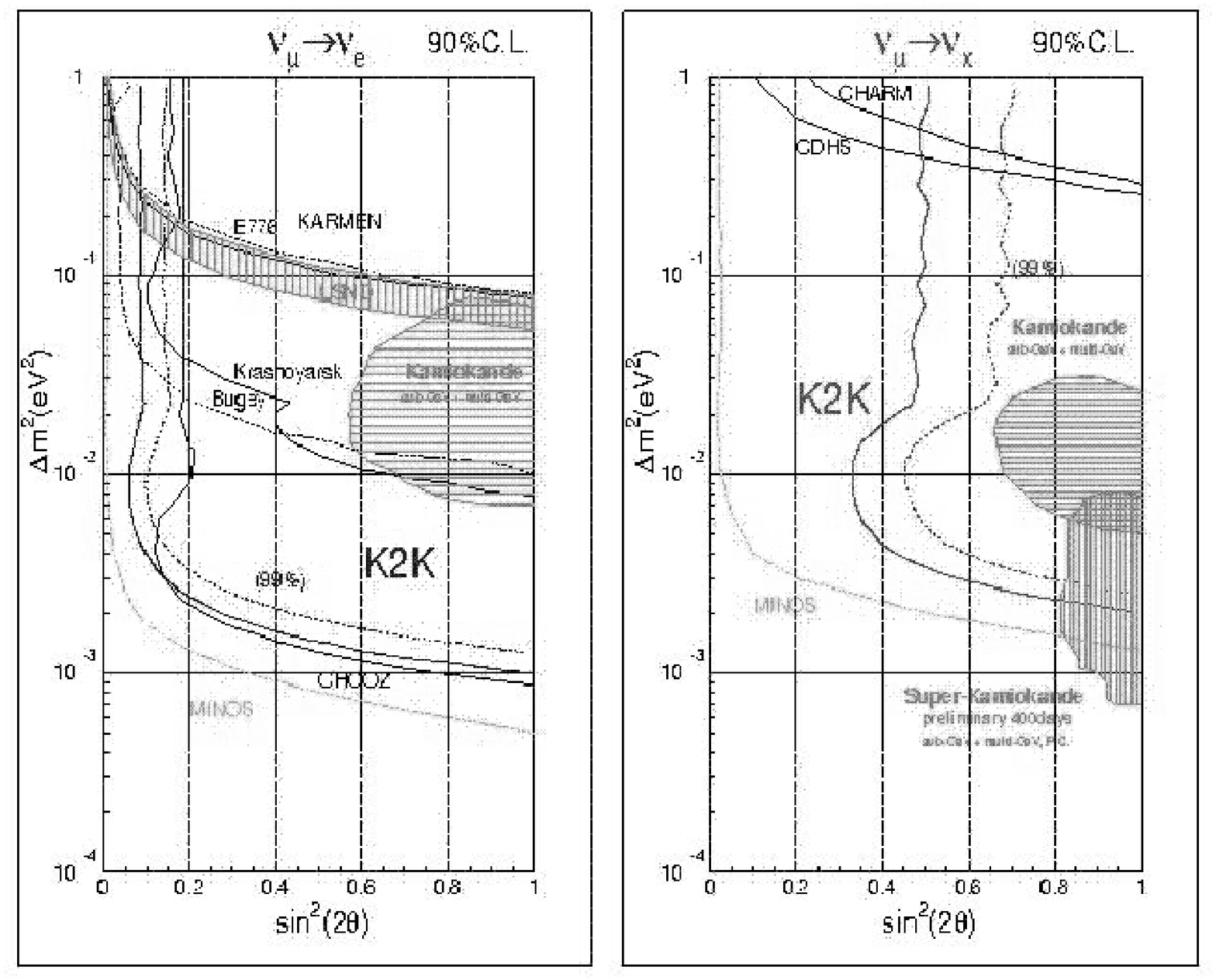}
  \end{center}
\label{fig:k2k}
\caption{Exclusion plot for $\nu_\mu$ disappearance after 3 years of 
operation of the k2k long-baseline project}
\end{figure}

The Japanese project is the most advanced of the three, since a successful
proton extraction was performed in the beginning of March and neutrino
physics is expected to start very soon. A close detector will be placed
close to the neutrino production for studies of flux and beam profile;
the far detector is SuperKamiokande. As can be seen in figure 
\ref{fig:nuspec}, the K2K beam has lower energy spectrum with respect to 
its competitors, covering mainly the atmospheric neutrino region. The mean
neutrino energy will be about 1.4 GeV, below $\tau$ production threshold,
so only $\nu_\mu$ disappearance can be performed. After $10^{20}$ protons
on target, corresponding to about 3 years of data taking, $\nu_\mu\to
\nu_x$ oscillations can be tested down to $2\times 10^{-3} eV^2$ (see
figure \ref{fig:k2k}), thus not completely covering the region indicated
by the atmospherics.\par
\subsection{High-energy beams}
The American and European projects share more similarities, and both are
expected to start in some years from now. A detailed comparison of the
two projects (table \ref{tab:lb}) shows that the larger number of protons
on target available in the American project is compensated by the fact that
the protons from the CERN SPS are more energetic, therefore the discovery
reach of the two beams is quite similar, especially if the SPS is run in
dedicated mode (indicated by an asterisk).
\begin{table}[tbh]
\begin{center}
\begin{tabular}{|l|l|l|}\hline
Parameter&NGS&NuMi\\ \hline
$E_p$ (GeV)&400&120\\
POT/year&$4.0\times 10^{19} (\times 2)^*$&$3.7\times 10^{20}$\\
$<E_\nu>$ (GeV)&26.7&17.6\\
$\nu_\mu$ CC/POT/ton&$4.7\times 10^{-20}$&$1.0\times 10^{-20}$\\
$\nu_\mu$ CC/y/ton&$1.9 (3.8)^*$&3.7\\ \hline
\end{tabular}
\end{center}
\label{tab:lb}
\caption{A comparison between the European and American long-baseline
projects. The asterisk indicates a dedicated SPS run.}
\end{table}
\subsection{NuMi program}
\begin{figure}[tbh]
  \begin{center}
   \includegraphics[width=11cm,bb=100 350 500 640,clip]{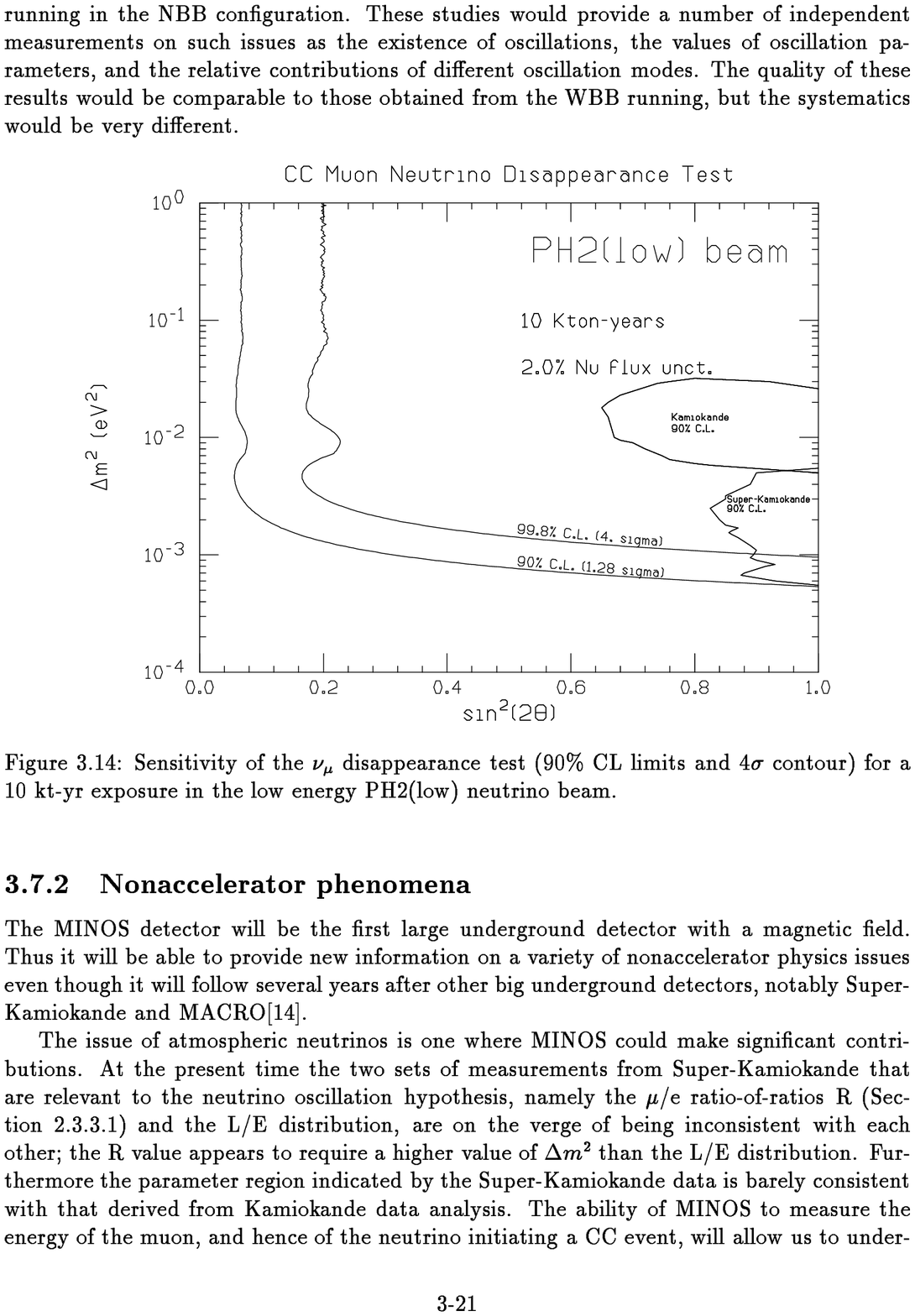}
  \end{center}
\label{fig:minoslim}
\caption{Exclusion plot for $\nu_\mu$ disappearance using the low-energy
NuMi beam.}
\end{figure}

The American long-baseline project foresees the production of neutrinos
with average energy of about 17.6 GeV from the Fermilab Main Injector,
and the detection in a 5.4 kton magnetized iron plate detector (MINOS 
\cite{minos}) placed
in the Soudan mine, in Minnesota. The total event rate will be as high as
$2\times 10^4$ charged current events per year, but the characteristics
of the detector do not allow a direct identification of the $\tau$.
Therefore only $\nu_\mu$ disappearance is possible, and the oscillation
parameters are measured via the ratio of neutral over charged current
events as a function of energy. At a given distance of 734 km, the first 
maximum of $\nu_\mu$ oscillation probability is given by $E_d= 594 GeV
\times \Delta m^2$, corresponding to 2.1 GeV for $\Delta m^2=3.5\times
10^{-3}$. This means that an optical configuration at which most of the
neutrino flux is at low value has to be chosen. In this case, a good
coverage of the atmospheric region can be achieved (figure 
\ref{fig:minoslim}).

\subsection{NGS program}
\begin{figure}[tbh]
  \begin{center}
   \includegraphics[width=10cm,bb=100 250 567 570,clip]{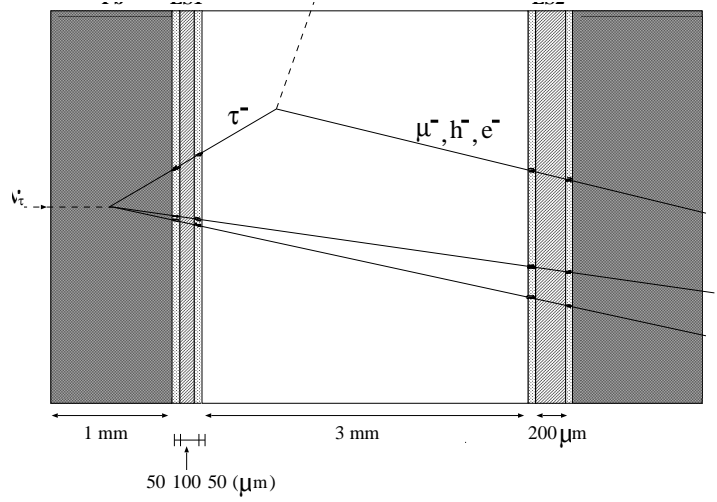}
  \end{center}
\label{fig:opera}
\caption{The basic structure of the OPERA detector. Neutrinos mainly
interact in the glass, producingsecondary particles traced in the
emulsions. In the case of $\tau$ production, they will decay in the air
gap, producing a large impact parameter in the second emulsion layer.}
\end{figure}

The main point of the European long-baseline program is that it is focused
on the direct detection of the $\tau$, directly derived from the experience
of $\tau$ search in Chorus and Nomad. Two complementary approaches are
possible:\begin{itemize}
\item topological identification of $\tau$ decays (kink search)
\item exploit particular kinematics of $\tau$ decays
\end{itemize}
The first approach is followed by the OPERA proposal \cite{opera}.
The homogeneous emulsion technique to search for $\tau$ production was
successfully tested and operated in CHORUS; since the neutrino flux at
Gran Sasso is much smaller, more mass is needed to have reasonable rates,
and a sandwich of a glass target and emulsions sheets used for tracking
is used, to achieve a final mass of 750 tons. Each lead-emulsion layer
is separated by their neighbors by a 3 mm air gap (see figure 
\ref{fig:opera}), so that most of the neutrinos interact in the glass,
and in case of oscillation the $\tau$ can decay in the air gap, producing
a kink visible as a large impact parameter, measured in the emulsion
layers, that have a detection granularity O($\mu$m).
\par
This technique should ensure very low background levels (about 0.1 
events/year), for a 90\% sensitivity to values of $\Delta m^2$ larger
than $1.5\times 10^{-3}$ for $\nu_\mu\to\nu_\tau$ oscillations with maximal
mixing.\par
\begin{figure}[tbh]
 \begin{minipage}{.45\linewidth}
  \begin{center}\mbox{
   \includegraphics[width=7cm]{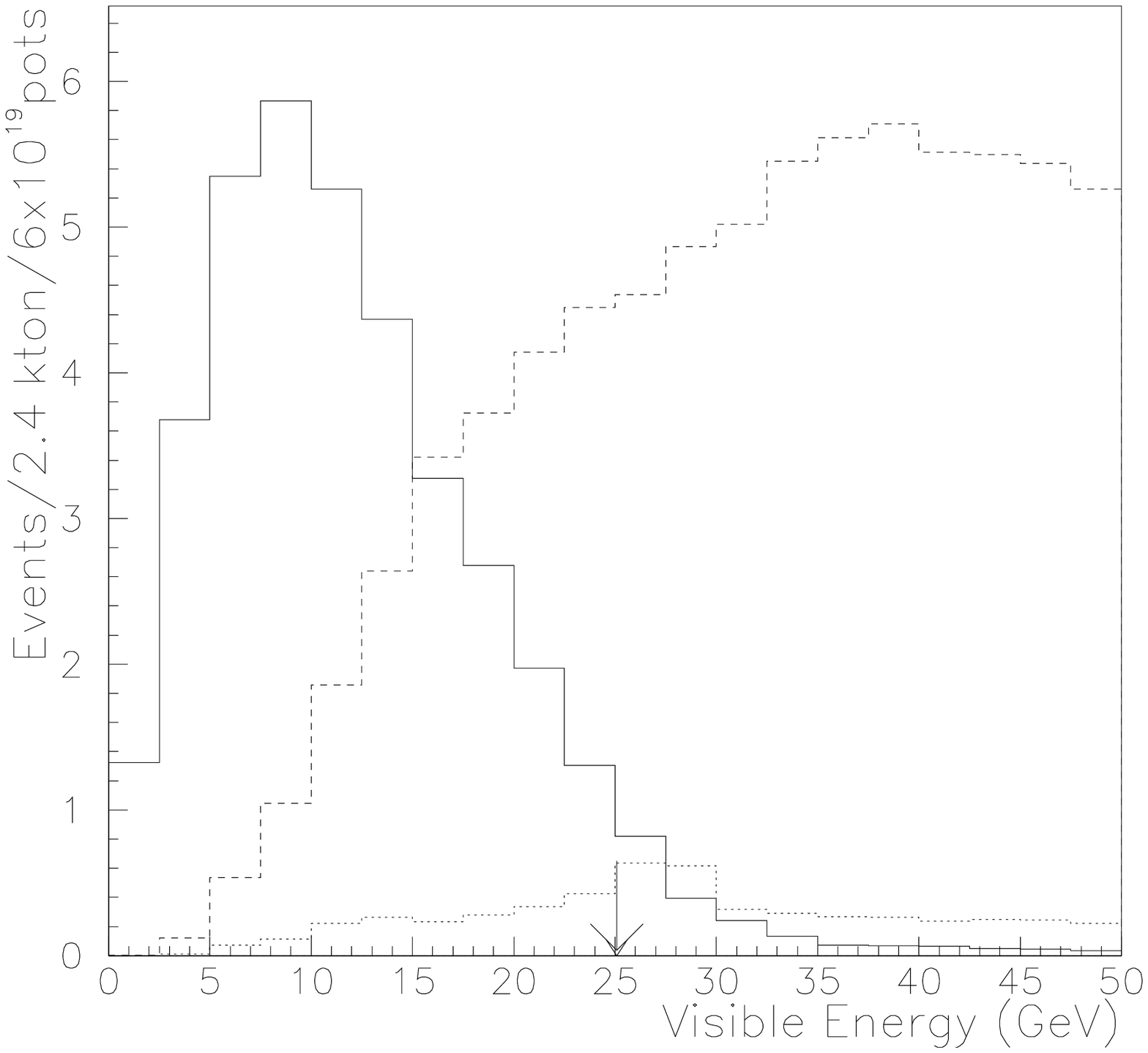}
}
  \end{center}
 \end{minipage}
 \begin{minipage}{.45\linewidth}
  \begin{center}\mbox{
   \includegraphics[width=7cm]{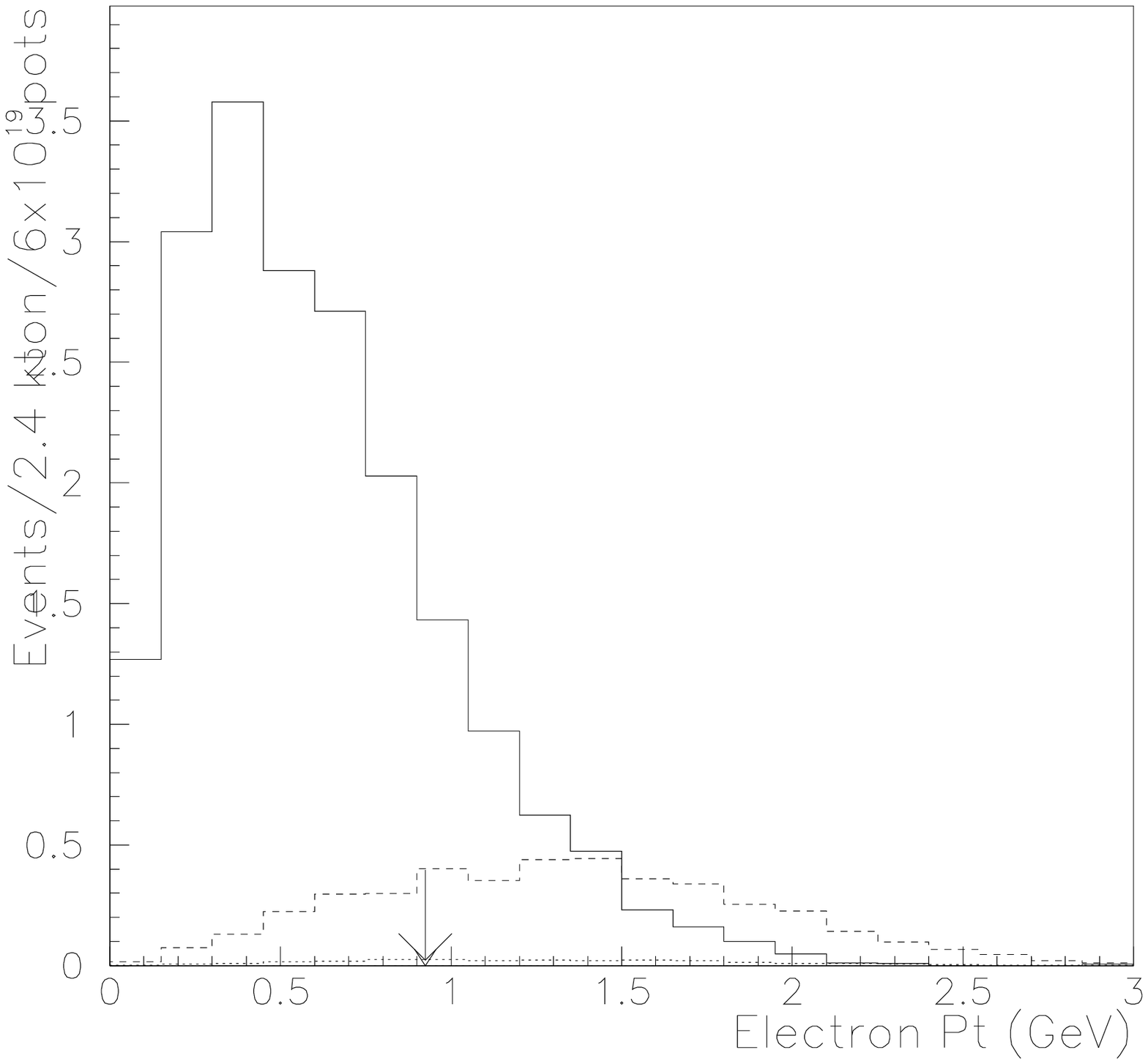}}
  \end{center}
 \end{minipage}
\label{fig:icatau}
\caption{Kinematica variables used in the $\tau\to e$ analysis in ICARUS}
\end{figure}

Another approach is to search for $\tau$ production using kinematic 
criteria, exploiting the particular characteristics of $\tau$ decays, as
it is foreseen by the ICARUS collaboration. Several $\tau$ decay channels
will be used, the $\tau\to e$ one being the cleanest. In this case the
main background is coming from $\nu_e$ charged-current interactions,
which are on average more energetic and more balanced. As can be seen from
figures \ref{fig:icatau}, cutting on few kinematical variables largely
improves the signal over background ratio, reaching sensitivity levels
comparable to those of OPERA for a 2.4 kton detector.
\par
\section{$\nu_\mu\to\nu_e$ oscillations}
The LSND experiment at Los Alamos \cite{lsnd} is so far the only 
accelerator neutrino detector to claim for positive evidence for neutrino
oscillations. The detector is composed of 167 tons of mineral oil and
scintillator, read by 1220 PMTs. The neutrinos are produced by Decay At
Rest (DAR) or In Flight (DIF) of muons from the Los Alamos Meson Facility,
and their mean energy is around 50 MeV, for a distance between production
and detection of about 30 m.\par
\begin{figure}[tbh]
  \begin{center}
   \includegraphics[width=6cm,clip,bb=0 100 512 720,clip]{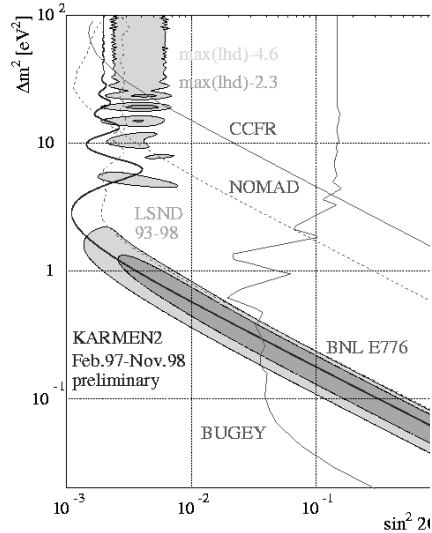}
  \end{center}
\label{fig:karmen}
\vspace{2cm}
\caption{Present $\nu_\mu\to\nu_e$ oscillation scenario. The KARMEN
result does not fully cover the allowed parameter space of the LSND claim.}
\end{figure}

In particular, LSND reports positive evidence for 
$\bar{\nu}_\mu\to\bar{\nu}_e$ oscillations in the following channels:
\begin{itemize}
\item $\bar{\nu_e} p\to e^+ n$, $n p\to d \gamma $(2.2 MeV) (DAR)\par
a $e^+$ tag comes from Cherenkov and scintillation light, in delayed
(180 $\mu s$) coincidence with a neutron tag 
\item $\nu_e C\to e^- X$ (DIF)\par
$e^+$ tag from Cherenkov and scintillation
\item $\nu_e ^{12}C\to e^- +^{12}N_{gs}, ^{12}N_{gs} \beta$ decay (DIF)\par
double, correlated $e^-$ $e^+$ tag
\end{itemize}
Since 1997 the Karmen2 experiment \cite{karmen} is running at RAL to 
verify the LSND
claim. They expect a total background of 7.8 events, mainly coming from
cosmic rays (1.7 ev.), $\nu_e$ in $\mu$ decays (2.6), accidentals (2.0)
and $\bar{\nu}_e$ beam contamination (1.5), and 8 candidates are observed.
As can be seen from figure \ref{fig:karmen}, only a part of the LSND
allowed region can be covered, and even more statistics will not fully
verify the LSND result.\par
\vspace{2cm}
\par
\begin{figure}[tbh]
  \begin{center}
   \includegraphics[width=8cm,clip,bb=115 290 525 680]{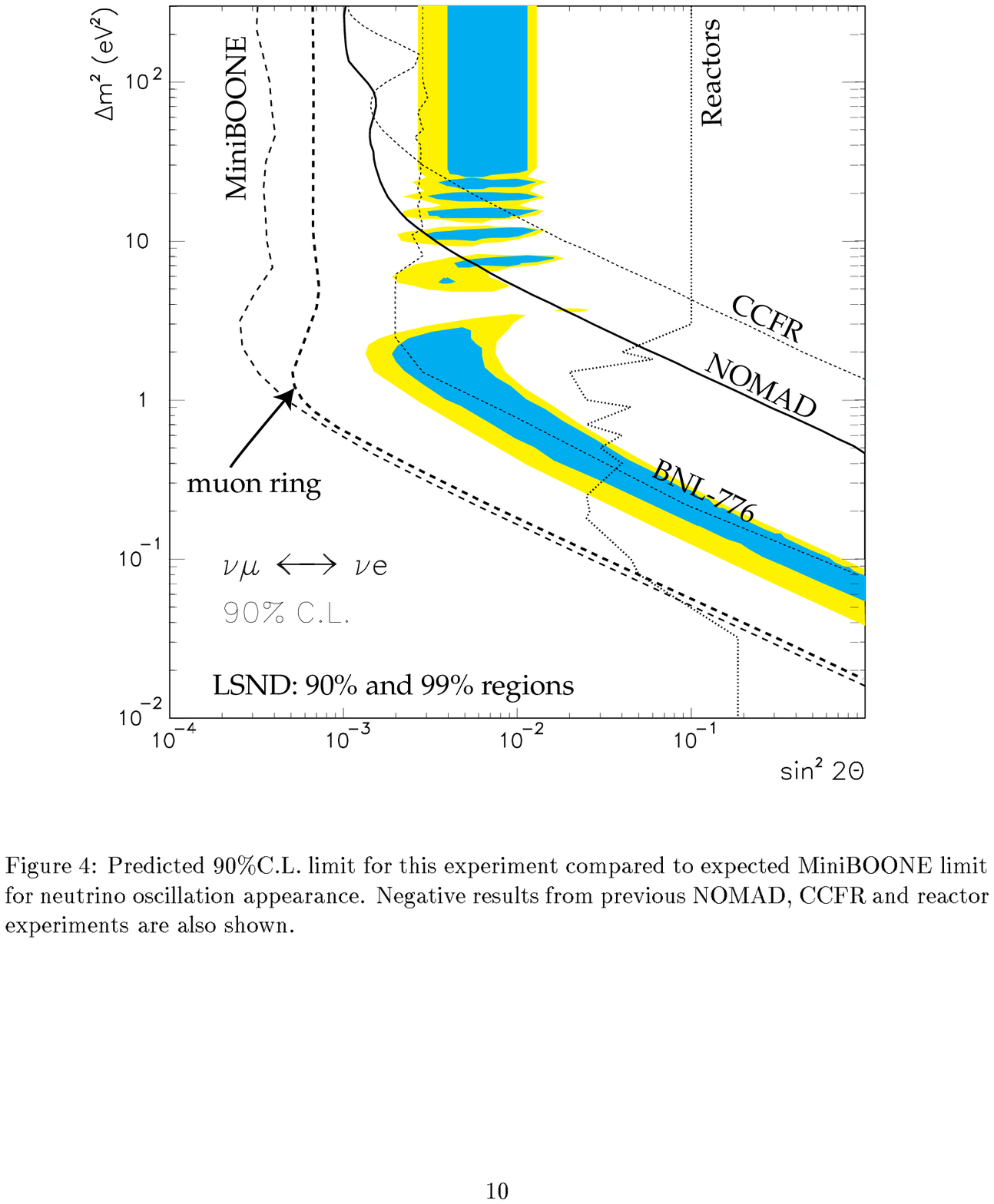}
  \end{center}
\label{fig:lsndprop}
\caption{Proposed experiments could fully cover the LSND parameter space
for $\nu_\mu\to\nu_e$ oscillations.}
\end{figure}
\par
Other new experiments are planned to verify $\nu_\mu\to\nu_e$ oscillations.
\par
The MiniBOONE \cite{mboone} experiment at the Fermilab booster has been 
approved, and is
expected to start taking data in year 2001. Its strategy is to look for
a large oscillation signal (about 1000 events in the LSND region)
over a large background (about 3000 events, out of which 1700 from beam 
$\nu_e$, 1200 from misidentification). From the purely statistical point of
view the significance of this approach is quite large, but some concern
exist on the control of the background ($\pi^0$ production, beam fluxes 
etc.)\par
The LoI-216 \cite{loi216} proposal at CERN is considering the use of the
existing low-energy neutrino beam from the PS (old Gargamelle line).
Neutrinos are sent to three identical modules, the first acting as a
near location at L=130 m and the other two as a far location L=885 m.
Each of the three modules will consist of a tracking calorimeter plus a
tail- and muon-catcher, for a total mass of $3\times 130$ tons.
The aim is to look for a variation of the $N_e/N_\mu$ ratio between the
two locations, in particular a signal is observed if
$(N_e/N_\mu)_{FAR}-(N_e/N_\mu)_{CLOSE}>0$\par
Using the existing CERN WANF beam, with $<E_\nu>\approx$ 27 GeV, a detector
placed on the Jura mountains (L=17 Km) \cite{jura} would be able to fully 
explore the
LSND region for $\nu_\mu\to\nu_e$ oscillations. The particularity of this
location is that with techniques similar to those already discussed for the
long-baseline case, a $\tau$ search is possible, testing the hypothesis 
that the LSND effect is instead due to $\nu_\mu\to\nu_\tau$ oscillations.
\par
A possibility for testing $\nu_\mu\to\nu_e$ oscillations with almost zero
background is to use neutrinos from muon decays instead of the 
``traditional'' beams from pions\cite{sbm}. In a storage ring, negative 
muons would
have the following decay $\mu^-\to e^-\bar{\nu_e}\nu_\mu$ producing
neutrinos of different flavors and opposite elicity. These neutrinos
will interact in a detector producing $\mu^-$ and $e^+$ in case of no
oscillation, while in case of 
$\bar{\nu}_e\rightsquigarrow\bar{\nu}_\mu$ or $\nu_\mu\rightsquigarrow\nu_e$
oscillations, leptons of opposite sign are observed. A light detector
with charge identification placed at few kilometers from the neutrino
production would test the LSND claim after few years of running.\par
As can be seen in figure \ref{fig:lsndprop}, these proposals can
fully cover the LSND suggested region.
\section{Conclusions}
Many experimental results in neutrino physics are pointing towards evidence
for neutrino oscillation. In particular:\begin{itemize}
\item the results from atmospheric neutrinos need confirmation with
different techniques and more precision, i.e. using large detectors
\item more information of $\nu_\mu$ disappearance may come from the K2K
long-baseline project quite soon; $\tau$ appearance will however only be 
possible using high-energy long baseline beams
\item many experiments are proposed or in preparation to look for $\nu_\mu
\to\nu_e$ oscillations, with a sufficient sensitivity to test the LSND
claim
\end{itemize}
Many experiments, either approved (SuperKamiokande, ICARUS) or proposed
(NOE, Aquarich, NICE, OPERA) can play a major role in one or many of the
forementioned topics, and neutrino physics promises to be a very hot topic
also in the years to come.
\section*{Acknowledgments}
I would like to thank the conference organizers for having invited me to
give this talk. For help in finding information about the different
experiments, I thank P.Strolin, P.Migliozzi, P.Picchi, F.Pietropaolo, 
R.Santacesaria, T.Ypsilantis, L.Ludovici, P.Zucchelli.
Many thanks to Andr\`e Rubbia and Antonio Bueno for useful discussions
during the preparation of the talk, and to Cristina Morone for the cover
picture.

\end{document}